\documentclass[12pt]{article}
\usepackage{amsmath}
\usepackage{amssymb}
\usepackage{amsfonts}

\usepackage[nomath]{stix}
\usepackage{setspace}
\usepackage[T1]{fontenc}
\usepackage{times}
\usepackage{palatino} 
\usepackage{lmodern}
\usepackage[latin1]{inputenc}
\usepackage{epsfig}
\usepackage[english]{babel}
\usepackage{color}
\usepackage{graphicx}
\usepackage{dcolumn}
\usepackage{moreverb}
\usepackage{cite}

\usepackage{arydshln}
\usepackage{cite}
\usepackage{braket}

\usepackage{dsfont}
\usepackage{nicefrac}
\usepackage{graphicx}
  \usepackage{stmaryrd}

\usepackage{array}
\usepackage{esvect}
\usepackage{cite}

\usepackage{physics}
\usepackage{mathrsfs}
\usepackage{authblk}

\usepackage[all,cmtip]{xy}

\numberwithin{equation}{section}

\title{The Lie algebra of the lowest transitively differential group of degree three}
\author{Alfred Michel Grundland$^1$, Ian Marquette$^{2}$\\ 
{\small $^1$ Centre de Recherches Math\'ematiques, Universit\'e de Montr\'eal }\\ 
{\small Succ. Centre-Ville, CP6128, Montr\'eal (Qc) H3C 3J7 Canada }\\ 
{\small Department of Mathematics and Computer Science, }\\
{\small Universit\'e du Qu\'ebec, CP500, Trois-Rivieres (Qc) G9A 5H7 Canada } \\
{\small $^2$ School of Mathematics and Physics, The University of Queensland,} \\ 
{\small Brisbane, QLD 4072, Australia} 
}

\date{ }
\begin{document}
\baselineskip=22pt plus 1pt minus 1pt
\maketitle

\begin{abstract} 

We investigate the real Lie algebra of first-order differential operators with polynomial coefficients, which is subject to the following requirements. (1) The Lie algebra should admit a basis of differential operators with homogeneous polynomial coefficients of degree up to and including three. (2) The generator of the algebra must include the translation operators $\partial_k$ for all the variables $x_1$,...,$x_k$. (3) The Lie algebra is the smallest indecomposable Lie algebra satisfying (1) and (2).  It turns out to be a 39-dimensional Lie algebra in six variables ($k=6$) and the construction of this algebra is also the simplest possible case in the general construction of the Lie algebras of the transitively differential groups introduced by Guillemin and Sternberg in 1964 involving the coefficients of degree 3. Those algebras and various subalgebras have similarities with algebras related to different applications in physics such as those of the Schr\"odinger, Conformal and Galilei transformation groups with and without central extension. The paper is devoted to the presentation of the structure and different decompositions of the Lie algebra under investigation. It is also devoted to the presentation of relevant Lie subalgebras and the construction of their Casimir invariants using different methods. We will rely, in particular, on  differential operator realizations, symbolic computation packages, the Berezin bracket and virtual copies of the Lie algebras. 
\end{abstract}

\\
email: grundlan@crm.umontreal.ca,i.marquette@uq.edu.au
\bigskip
\noindent
%
%

\newpage

\section{Introduction}

The problem of the classification of algebras of vector fields has been central in physics. It was originally formulated by Sophus Lie and pursed over the year \cite{kam1,kam2, kam3, pos1}. 
Lie algebras and their differential operators realizations play an important role in the theory of solvable, quasi-exactly solvable and integrable systems \cite{tur1,tur2}. Among Lie algebras of interest for physical application, is the class of transitively differential groups and their Lie algebras  \cite{gui64, post68, kan94}. A Lie group G of transformations of a manifold with coordinates $x_1$,....,$x_n$ is called a transitively differential group (TD-group) of degree m if its Lie algebra g fulfills the following conditions:

1. There exists a basis of infinitesimal operators 
\[ f^j(x_1,x_2,...,x_n)\partial_j\]
whose coefficients are homogeneous forms of degree $k \leq m$; $\max \quad k=n$.
\\
2. The Lie algebra $g$ contains all operators of the form
\[ \partial_j ,\quad j=1,2,...n \]
\\
3. The Lie algebra $g$ is maximal i.e. g is not properly contained in any other Lie algebra satisfying (1) and (2).

The definition shows why these groups are called transitively differential: translations in all space directions are contained in it by definition. From the physics point of view, this corresponds to a homogenous space over which the algebra is modeled. A prominent example is the conformal group, which is of degree 2, so that the corresponding Lie algebra $g$ consists of differential operators with polynomial coefficients of degree less than or equal to two. Whereas Lie algebras of TD-groups of degree less than or equal to 2 are often used in physics (other examples are e.g. the group of affine transformations (of degree 1) and the group of projective transformations (of degree 2)), only little attention has been paid to Lie algebras of TD-groups of degree 3 \cite{gui64,post68,kan94}. More generally those algebras are not semisimple. Their classification and the study of their invariants \cite{Sn05,Sn14} is difficult. Some classes, which are similar to the non-semisimple algebra studied in this paper, are among the ones that can be applied in different areas of physics, such as the Schr\"odinger algebra describing the maximal dynamical symmetries of the Schr\"odinger equation \cite{Ba68,Ni72,Pe77,Do97,Ai02}. The algebra is connected with a cubic Casimir invariant and a generalization of hypergeometric and nontrivial induced representations. Various generalizations, such as the conformal Galilei algebra with and without a central extension, also play a central role \cite{Ai10,Ai11, Ai12,Ai13,An14}. The construction of a Casimir invariant for a simple Lie algebra \cite{Po66,Ok77,Ab75,Be81} has been accomplished. The problem of the construction of Casimir invariants for non-semisimple Lie algebras is difficult. However, various methods have been introduced to address this issue \cite{Ca03, Ca09, Qu88, Sn14, fad18, fad19, rut19}. Further results on enveloping algebras and sets of polynomials related to commutant and integrable models have been found in recent years \cite{rut22}.

Here, the Lie algebra $W$ of a transitively differential group of degree 3,  as introduced in \cite{kan94}, is investigated in detail.

We present a decomposition with an unbalanced Z-grading. We also present different subalgebras and their Casimir invariants, some results on central extensions and a further understanding of the enveloping algebra. The Lie algebra of the lowest transitively differential group of degree three is then generated by the following 39 generators $\{A_1,...,C_7\}$

\begin{equation}
  A_1= \partial_{x_1},\quad A_2= \partial_{x_2},\quad  A_3=\partial_{x_3},\quad A_4= \partial_{x_4} ,\quad A_5=\partial_{x_5},\quad A_6= \partial_{x_6} \label{real}
\end{equation} 
\[ L_1= x_1 \partial_{x_6},\quad L_2= x_2 \partial_{x_6},\quad L_3=  x_3 \partial_{x_6},\quad L_4= x_4 \partial_{x_6},\quad L_5= x_5 \partial_{x_6},\quad L_6= x_6 \partial_{x_6}  \]
\[ L_7=  x_1 \partial_{x_3},\quad L_8 =  x_3 \partial_{x_3},\quad  L_9 =  x_1 \partial_{x_4},\quad L_{10} =  x_2 \partial_{x_4},\quad  L_{11} =  x_1 \partial_{x_5},\quad L_{12} =  x_2 \partial_{x_5}\]
\[ L_{13}= x_1 \partial_{x_1} +2 x_3 \partial_{x_3} + x_4 \partial_{x_4} ,\quad L_{14}= x_2 \partial_{x_1} +2 x_4 \partial_{x_3} + x_5 \partial_{x_4} ,\quad L_{15}= x_1 \partial_{x_2} + x_3 \partial_{x_4} + 2x_4 \partial_{x_5} \]
\[ L_{16}= x_2 \partial_{x_2} + x_4 \partial_{x_4} + 2 x_5 \partial_{x_5} ,\quad  L_{17}= x_3 \partial_{x_3} + x_4 \partial_{x_4} + x_5 \partial_{x_5} ,\quad  B_1= x_1^2 \partial_{x_6} , B_2= x_1 x_2 \partial_{x_6} , \]
\[  B_3= x_2^2 \partial_{x_6},\quad B_4 = (x_1 x_5 - x_2 x_4) \partial_{x_6} ,\quad  B_5 = (x_2 x_3 - x_1 x_4 ) \partial_{x_6} ,\quad  B_6= \frac{1}{2} x_1 x_2 \partial_{x_4} +x_2^2 \partial_{x_5} \]
\[ B_7= x_1^2 \partial_{x_4} + 2 x_1 x_2 \partial_{x_5} ,\quad B_8= 2 x_1 x_2 \partial_{x_3} + x_2^2 \partial_{x_4},\quad  B_9= x_1^2 \partial_{x_3} + \frac{1}{2} x_1 x_2 \partial_{x_4}\]
\[ C_1= x_1^3 \partial_{x_6},\quad  C_2= x_1^2 x_2 \partial_{x_6} ,\quad  C_3= x_1 x_2^2 \partial_{x_6},\quad  C_4= x_2^3 \partial_{x_6} \]
\[ C_5= x_1^3 \partial_{x_3} + x_1^2 x_2 \partial_{x_4} +x_1 x_2^2 \partial_{x_5}, \quad C_6= x_1^2  \partial_{x_3} + x_1 x_2^2  \partial_{x_4} + x_2^3  \partial_{x_5}\]
\[ C_7= (x_1^2 x_5 + x_2^2 x_3 -2 x_1 x_2 x_4)  \partial_{x_6} \]

They are expressed as $X_i=\sum_j^6 Q_j^(i) \partial_{x_j}$ where $Q_j^(i)$ are polynomials in the 6 variables. Then, the generators $A_i$ for $i=1,...,6$ are generators where the $Q_j^(i)$ are of degree 0, $L_i$ for $i=1,...,17$ are generators where the $Q_j^(i)$ are of degree 1, $B_i$ for $i=1,...,9$ are generators where the $Q_j^(i)$ are of degree 2 and $C_i$ for $i=1,...,7$ are generators where the $Q_j^(i)$ are of degree 3.

We then obtain the commutation relations of the generators in this basis (\ref{real}). We only present non-vanishing commutators (see Appendix A). They will be used in the next section to obtain the decomposition and different subalgebras of interest.

\section{Decomposition of the Lie algebra W}

The Lie algebra W is decomposed as $W= U_{-1} \oplus U_0  \oplus U_1  \oplus U_2$ where $U_{0}$,  $U_1$,  $U_{-2}$ and $U_{-1}$ are given in following way, (1) $U_{-1}$ is generated by $A_i$ where $i=1,..,6$, (2) $U_{0}$ is  generated by $L_j$ where $j=7,..,17$, (3) $U_{1}$ is generated by $B_k$ where $k=1,..,9$ and (4) $U_{2}$ is generated by $C_s$ where $s=1,..,7$. This is an unbalanced $Z$-graded Lie algebra as the grading is given by $\{-1,0,1,2\}$. Usually, graded or Z graded Lie algebras which appear in the context of physics admit a grading such as $\{-1,0,1\}$ or $\{-2,-1,0,1,2\}$ as is the case for Schr\"odinger and conformal algebras or even different conformal Galilean algebras with central extensions \cite{Ai10,Ai11,Ai12,Ai13,An12,An14}. 

The dimensions of the different subspaces are

\[ dim(U_{-1})=6 ,\quad dim(U_0)=17 ,\quad dim(U_1)=9 ,\quad dim(U_2)=7 \] 

Using the formula from Appendix A, we can verify explicitly that
\begin{equation}
 [U_j, U_p]= U_{j+p}
\end{equation}
However, if $j+p \neq \{-1,0,1,2\}$ then $[U_j,U_p]=0$. The algebra admits different subalgebras, among them $sl(2)=\{J_2,J_0,J_{-2}\}$ such as
\begin{equation}
J_2= L_{14} ,\quad J_0 = L_{16}-L_{13} ,\quad J_{-2}=L_{15} 
\end{equation}
where
\begin{equation}
[J_0,J_2]= 2 J_2 ,\quad [J_0,J_{-2}]=-2 J_{-2} ,\quad [J_2,J_{-2}]=-2J_{-2} 
\end{equation}
which includes Abelian and non-Abelian radicals with the following decomposition
\begin{equation}
V_2= sl_2 \niplus U_2 ,\quad V_{12}= sl_2 \niplus U_{1} \oplus U_{2} ,\quad V_{-1}= sl_2 \niplus U_{-1} ,\quad V_{0}= sl_2 \niplus U_{0} 
\end{equation}
which has similarities with the non-centrally extended Schr\"odinger algebra.
We will give further details on the structure of those subalgebras and this will provide insight into a change of basis for this algebra. The notation includes the following: $S$ indicates a singlet, $D$ a doublet, $T$ a triplet and $Q$ a quadruplet. The upper index indicates the subspace it belongs to in terms of the grading i.e. $-1,0,1,2$, and the value in the parenthesis is the eigenvalue relative to $J_0$, which is the Cartan subalgebra from the $sl(2)$ algebra, i.e. the semisimple part of the Levi decomposition of the Lie algebra $W$. If, in a same subspace, there is more than one singlet, doublet, triplet or quadruplet, we distinguish the different element with a lower index.

For example, $T_2^1(-2)$ would correspond to the second triplet in the subspace associated with the grading $U_1$ and be the element with eigenvalue $-2$. We have organized the multiplets in such a way that the eigenvalues of singlets are $0$, of doublets are $-1,1$, of triplets are $-2,0,2$ and of quadruplets are $-3,-1,1,3$. The $U_{-1}$ algebra has a singlet, a doublet and a triplet
\[ S^{-1}(0)= A_6 ,\quad  D^{-1}(1)= A_1 , D^{-1}(-1)=A_2 \]
\[ T^{-1}(2)=A_3 , T^{-1}(0)=A_4 , T^{-1}(-2)=A_5, \]
while for the $U_{2}$ algebra, we have a quadruplet, a doublet and a singlet
\[ Q^{2}(3)= C_4 , Q^{2}(1)=C_3 , Q^2(-1)=C_2 , Q^2(-3)=C_1 \]
\[ D^{2}(1)=C_6, D^2(-1)=C_5 ,\quad S^{2}(0)=C_7. \]
For the $U_{1}$ algebra, we have two triplets, a doublet and a singlet
\[ T_1^1(2)= B_3 , T_1^1(0)= B_2, T_1^1(-2) = B1 \]
\[ D^1(1) = B_4, D^1(-1) = B_5 ,\quad S^1(0)= B_6+ B_9\]
\[ T^1_2(2)= B_8, T^1_2(0)= B_9-B_6 , T_2^1(-2) = B_7 \]
and for the last component in the grading, the $U_{0}$ algebra, we have a quadruplet, a doublet and two singlets 
\[ Q^0(3)= L_{11}, Q^0(1)= L_{12}-L_{9}, Q^0(-1)= 2(L_7-L_{10}), Q^0(-3) = 6 L_8 \]
\[ D^0_1(1)= 2 L_{12} + L_9 , D^0_1(-1)= L_{10}+2 L_7 ,\quad S^0_1(0) =L_{13} +L_{16}, S^0_2(0)= L_{17} \]
\[ S^0_3(0) =L_6,\quad  D^0_2(-1)=L_1,\quad D^0_2(1)=L_2 ,\quad  T^0(-2)= L_3,\quad T^0(0)=L_4,\quad T^0(2)= L_5 \]

It is possible to use those singlets, doublets, triplets and quadruplets to split every subspace $U_i$ into subspaces which are irreducible with respect to $sl(2)$. We have the following decomposition

\[ U_0= sl(2) \oplus 3 S_0 \oplus 2 D_0 \oplus T_0 \oplus Q_0 \]

\[ U_{-1}= S_{-1} \oplus D_{-1} \oplus T_{-1} ,\quad U_1= S_1 \oplus D_1 \oplus 2 T_1 ,\quad U_2= S_2 \oplus D_2 \oplus Q_2 \]

Let us also define $\tilde{U}_0= 3 S_0 \oplus 2 D_0 \oplus T_0 \oplus Q_0$.
The Levi decomposition is then given by

\[ W= sl(2) \niplus R \]

where $R$ is given by $\{U_{-1},\tilde{U}_0,U_1,U_2\}$ and is of dimension 36.

\section{Casimir invariant of the subalgebras of W}

Casimir invariants for simple Lie algebras are well known and there are different ways of expressing them, such as using the Killing form and the determinantal formula. On the other hand, non-semisimple Lie algebras do not have such a direct construction. However, different methods are known for constructing Casimir invariants for those non-semisimple Lie algebras \cite{Ca03, Ca09, Qu88, Sn14,fad18, fad19, rut19}. They are, however, difficult to apply in practice and here, in the context of this Lie algebra of dimension 39, even for subalgebras, they would lead to difficult computational problems. 

\begin{itemize}

\item[(1)] Among the methods we plan to rely on in this paper is the method of virtual copies of Lie algebras \cite{Ca09,rut19}, which consists of constructing a map from the enveloping algebra of a non-semisimple Lie algebra to a semisimple Lie algebra and then using various methods, such as the Gelfand determinantal and symmetrization map, to get an explicit formula for the Casimir invariants. Those methods do not apply to non-semisimple Lie algebras with Abelian radicals, so other methods need to be applied. In some cases, the virtual copy in the enveloping algebra does not provide enough Casimir. This happens if the dimension of the radical reaches certain dimensions compared with the semisimple part of the Levi decomposition.

\item[]

\item[(2)] The infinitesimal method \cite{Ca09,Sn14} for solving partial differential equations consists of using the Poisson-Lie bracket setting on the variables of the dual space that are commutative and then solving the linear systems using the method of characteristics. The symmetrization map is applied to the solution written in terms of the variables of the dual space in order to obtain the Casimir invariants of the corresponding Lie algebra in terms of the generators. This method relies, not only on the correspondence between the Lie and the Poisson-Lie algebra, but also on a one-to-one correspondence between their enveloping algebras through the symmetrization map. This method is, in some ways, the most general as it goes beyond polynomial invariants, and allows us to find even non-polynomial solutions, which are then viewed as generalized Casimir invariants. This method is, however, difficult to apply in practice when the dimension increases as it relies on the method of characteristics to solve the systems of partial differential equations.

\item[]

\item[(3)] A recently introduced method, which consists of constructing Casimir invariants using a realization \cite{fad18,fad19}, can be applied in our case as we have started with an explicit realization. This can offer computational advantages but provides a larger set of polynomials which are not Casimir invariants as they commute only in the realization and not as polynomials in the enveloping algebra. They might be interesting in their own right and in regards to the theory of representations and related special functions of that specific realization. A root system is not available in the context of non-semisimple Lie algebras, but other ideas such as the dimensional analysis approach, automorphism and other structural properties can be used to facilitate those calculations \cite{fad18,fad19}. In the context of our Lie algebras of dimension 39, we have already identified a grading which can be used as well as eigenvalues relative to different generators. This can provide a reduction in terms of the types of monomials that are allowed in the Casimir invariant.

\item[]

\item[(4)] Our approach consists of using a symbolic computation package such as the NCalgebra package \cite{ncalg}, which allows us to manipulate non-commutative variables and polynomials of those variables in order to construct Casimir invariants. Many packages were introduced, among them the NCalgebra package,  which allows us to manipulate Lie algebras and applies the reordering of terms according to of a PBW basis. This package was used in a series of papers \cite{rut22,dan22}. The approach consists of constructing Casimir from arbitrary linear combinations of admitted monomials in the generators, getting linear equations with unknown coefficients, solving the systems and obtaining the linearly independent Casimir for every remaining undetermined coefficient. Further calculations need to be performed to then find an algebraically independent set. This method has been particularly useful for getting elements beyond Casimir invariants. Those elements consist of commutants or partial Casimir invariants \cite{rut22,dan22}. This method could be seen as a direct approach for the obtention of polynomial invariants, and it is difficult in practice to apply it when the degrees of the Casimir invariants are increasing as well as the dimension of the underlying Lie algebra. Again, other properties can allow us to reduce the allowed monomials via structural properties of the Lie algebras.

\item[]

\item[(5)] A related approach consists in applying the previous ideas of method 4 i.e. considering arbitrary linear combinations of all allowed monomials, but instead using the Berezin bracket \cite{bo09,ki76}, and then constructing general polynomials up to a given degree and applying the symmetrization map to any Casimir obtained in this way. This can be less involved than the other approach from a computational point of view and it also allows us to get access to elements in the enveloping algebra beyond the Casimir invariants \cite{mar22}. This approach offers a compromise between method 2 that relies on solving systems of PDEs and method 4 which consists of using non-commutative variables. In practice, this method allow us to get further insight than those other methods. 

Other approaches can also be defined, for example, those based on coadjoint representations and orbits. These methods have been introduced. The algebra obtained in this paper offers the possibility of applying those methods to get insight into the Casimir invariants.

\end{itemize}

The general problem of looking at Casimir invariants has been investigated with different approaches, in particular methods 4 and 5, taking the ordering $\{A_1,...,C_7\}=\{Y_1,...,Y_{39}\}$. We then look for general polynomials of the form

\begin{equation}
K=\sum_{S \leq N} \sigma_{\alpha_i} \prod_j^{39} Y_j^{\alpha_j}
\end{equation}
where $S=\sum_{k}^{39}\alpha_k$ and $\sigma(\alpha_i)= \sigma(\alpha_1, ... ,\alpha_{39})$. Second, third and fourth degree Casimir (i.e. N=2,3,4) has been found to be only trivial. The various subalgebras correspond to $sl(2)$ acting decomposable and indecomposable representations. This generalizes the case of the Schr\"odinger and conformal Galilean-type algebras (in the case where $d=1$, i.e. only one spatial dimension). This makes those subalgebras interesting in their own right from the perspective of invariants.

\subsection{Subalgebra $V_2= sl_2 \niplus U_2$}

Let us investigate the decomposition of the subalgebra $V_2= sl_2 \niplus U_2$. We can see that $C_7$ forms a singlet under the action of $sl_2$, $\{C_5,C_6\}$ forms a doublet under the action of $sl_2$ and $\{C_1,C_2,C_3,C_4\}$ forms a quadruplet under the action of $sl_2$. We can consider an alternative notation $S(0)=C_7$, $D(1)=C_6$, $D(-1)=C_5$, $Q(3)=C_4$, $Q(1)=C_3$, $Q(-1)=C_2$, $Q(-3)=C_1$. The non-vanishing commutation relations are

\begin{align}
&[J_0,J_2]= 2 J_2,\quad &[J_0, J_{-2}]=-2 J_{-2},\quad &[J_2,J_{-2}]=J_0 \\ \nonumber
&[J_0,C_1]=-3 C_1 ,\quad &[J_0,C_2]=- C_2,\quad &[ J_0,C_3]=C_3\\ \nonumber
&[J_0,C_4]=3 C_4,\quad &[J_0,C_5]= -C_5, \quad &[J_0,C_6]=C_6\\ \nonumber
&[J_0,C_7]=0,\quad &[J_2, C_1]=3 C_2,\quad &[J_2,C_2]=2 C_3\\ \nonumber
&[J_2,C_3]=C_4,\quad &[J_2,C_4]=0 ,\quad &[J_2, C_5]=C_6\\ \nonumber
&[J_2,C_6]=0,\quad &[J_2,C_7]=0,\quad &[J_{-2},C_1]=0 \\ \nonumber
&[J_{-2},C_2]=C_1,\quad &[J_{-2},C_3]=2 C_2,\quad &[J_{-2},C_4]=3 C_3\\ \nonumber
&[J_{-2},C_5]=0,\quad &[J_{-2},C_6]=C_5,\quad &[J_{-2},C_7]=0 \nonumber
\end{align}

As the problem is lower-dimensional, we start by applying the fourth approach consisting of using the PBW basis with the following ordering $\{ J_2,J_0,J_{-2},C_1,C_2,C_3,C_4,C_5,C_6,C_7 \}$ and we then look for a Casimir that takes the form

\[ K= \sum_{S \leq N}  \sigma_{\alpha_i} \prod_j^{10} X_j^{\alpha_j} \]

where $S= \sum_i^10 \alpha_i$ and

\[\sigma(\alpha_i)=\sigma(\alpha_1,\alpha_2,\alpha_3,\alpha_4,\alpha_5,\alpha_6,\alpha_7,\alpha_8,\alpha_9,\alpha_{10})\]

Using $N=1,2,3,4,5,6$, we are able to perform an extensive search for Casimir invariants. The problem of considering constraints arising from the conditions

\[  [X_i,K]=0\quad \forall \quad i=1,...,10. \]

Reordering the terms according to the PBW basis and selecting coefficients for those monomials in the generators of the algebra provides a set of linear equations for the coefficients $\sigma_{\alpha_i}$. Solving those equations, there remain 4 coefficients which provide the 4 linearly independent Casimir invariants that are also algebraically independent

\begin{align}
&K_1= C_7  \label{k1} \\
&K_2= C_3^2 C_2^2 - \frac{4}{3} C_3^3 C_1 - \frac{4}{3} C_4 C_2^3 + 2 C_4 C_3 C_2 C_1 - \frac{1}{3} C_4^2 C_1^2 \label{k2}\\
&K_3= C_5^2 C_3^2 - C_5^2 C_4 C_2 - C_6^2 C_2^2 + C_6^2 C_3 C_1 \label{k3} \\
&K_4= C_5^3 C_4 - 3 C_6 C_5^2 C_3 +3 C_6^2 C_5 C_2 - C_6^3 C_1  \label{k4}
\end{align}

Here, we notice that all the monomials in the four Casimir invariants have global degree 0 with respect to $J_0$ i.e. their eigenvalues are 0 when commuting with $J_0$. This implies that the search for monomials can be reduced by using the various properties of Casimir invariants, such as eigenvalues relative to $J_0$. We will now go to the Poisson-Lie algebra setting and apply a similar construction with commutative variables of the dual space to implement the method 5. We will then use the variables of the dual space and the infinitesimal method, which corresponds to what we denoted as method 2, to solve the systems of partial differential equations and obtain generalized Casimir invariants.

Taking into account that, we have a Lie algebra with generators $\{X_1,...,X_n\}$ and commutation relations $[X_i,X_j]=C^{k}_{ij} X_k$, where the $X_i's$ are realized by mean of the first-order differential operators, we consider the vector field

\begin{equation}
\hat{X_i}=c_{ij}^k x_k \frac{\partial}{\partial x_j}.\label{vecf}
\end{equation}

The Lie-Poisson bracket for the Lie algebra corresponding to the coordinates $\{x_1,...,x_n\}$ is

\begin{equation}
 \{x_i,x_j\}=C_{ij}^k x_k, 
\end{equation}

where $\{A,B\}$ is a Poisson bracket. Given smooth functions f and g $\in$ $C^{\infty}(g*)$, we have

\[ \{f,g\}= C^k_{ij} x_k \partial_{x_i} f \partial_{x_j}g, \]

which is equivalent to

\[ \{f,g\}= \sum_{i <j}( \frac{\partial f}{\partial x_i} \frac{\partial g}{\partial x_j}-  \frac{\partial f}{\partial x_j} \frac{\partial g}{\partial x_i}) \{x_i,x_j\} \]

This is connected with the Berezin bracket \cite{bo09}. We change the notation of the variables $x_i$ in order to denote commutative variables of the dual space of the Lie algebra

\[ J_0 = X_1,\quad J_2=X_2,\quad J_{-2}=X_3,\quad C_1=X_4, \quad C_2=X_5\]
\[ C_3=X_6, \quad C_4 = X_7,\quad C_5=X_8,\quad C_6=X_9,\quad C_7=X_{10} \]

and then consider the Poisson bracket for the coordinates of the dual space $x_1,...,x_{10}$

\begin{align}
&\{x_1,x_2\}=2 x_2,\quad &\{x_1,x_3\}=-2x_3, \quad  &\{x_1,x_4\}=-3 x_4,\quad &\{x_1,x_5\}=-x_5\\ \nonumber
&\{x_1,x_6\}=x_6,\quad &\{x_1,x_7\}=3 x_7,\quad &\{x_1,x_8\}=-x_8,\quad &\{x_1,x_9\}=x_9,\\ \nonumber
&\{x_1,x_{10}\}=0,\quad &\{x_2,x_3\}=x_1,\quad &\{x_2,x_4\}=3 x_5,\quad &\{x_2,x_5\}=2 x_6,\\ \nonumber
&\{x_2,x_6\}=x_7,\quad &\{x_3,x_5\}=x_4,\quad &\{x_3,x_6\}=2 x_5,\quad &\{x_3,x_7\}=3 x_6,\\ \nonumber
&\{x_2,x_8\}=x_9,\quad &\{x_2,x_8\}=x_9,\quad &\{x_3,x_5\}=x_4,\quad &\{x_3,x_6\}=2 x_5 ,\\ \nonumber
& \{x_3,x_7\}=3 x_6,\quad &\{x_3,x_9\}=x_8 & &\\ \nonumber
\end{align}

We, then consider an expansion in the commutative variables of the dual space

\[ \tilde{K}= \sum_{S \leq N}  \sigma_{\alpha_i} x_1^{\alpha_1} x_2^{\alpha_2} x_3^{\alpha_3} x_4^{\alpha_4} x_5^{\alpha_5} x_6^{\alpha_6} x_7^{\alpha_7} x_8^{\alpha_8} x_9^{\alpha_9} x_{10}^{\alpha_{10}} \]

with the constraints

\[  \{x_i,K\}=0 ,\quad \forall i=1,...,10 \]
which, again, provides an expression involving monomials ( now in terms of commutative variables ). Up to degree 4, we have 1001 unknown coefficients corresponding to each of the allowed monomials in the 10 variables ( 330 if 7 variables are related to the radical) and 878 equations. We obtain solutions in terms of the four remaining coefficients and give the following 4 Casimir invariants

\begin{align}
& \tilde{K}_1=x_{10} \label{k1t}\\
& \tilde{K}_2= -3 x_5^2 x_6^2 +4 x_4x_6^3 + 4 x_5^3 x_7 - 6 x_4 x_5 x_6 x_7 + x_4^2 x_7^2 \label{k2t}\\
& \tilde{K}_3= - x_6^2 x_8^2 + x_5 x_7 x_8^2 + x_5 x_6 x_8 x_9 - x_4 x_7 x_8 x_9 - x_5^2 x_9^2 + x_4 x_6 x_9^2 \label{k3t} \\
&\tilde{K}_4= -x_7 x_8^2 +3 x_6 x_8^2 x_9- 3 x_5 x_8 x_9^2 + x_4 x_9^3  \label{k4t}
\end{align}

In general, one would need to apply the symmetrization map

\[\Phi\left(x_{i_1}x_{i_2}\dots x_{i_s}\right)=\frac{1}{s!} \sum_{\sigma\in\Sigma_{s}} X_{i_{\sigma(1)}}X_{i_{\sigma(2)}}\dots X_{i_{\sigma(s)}},\]

Here, however all the Casimir invariants depend on the radical, which is Abelian, and there is no need to reorder the terms of the monomials involved in those expressions. Hence, the $\tilde{K}_i$ provide the $K_i$. We will now consider more general invariants $\bar{K}$ of the Lie algebra $V_2$. In particular, the generalized Casimir corresponds to the solution of the following system of partial differential equations
\[\hat{X}_i \bar{K} =0 ,\quad 1 \leq i \leq n \]
where $\bar{K}=\bar{K}(x_1,...,x_n)$. This reduces the problem to the solution of a system of PDEs. From $\{\bar{K},x_1\}=0$ to $\{\bar{K},x_{10}\}=0$, we obtain the following equation where ($\partial_i = \frac{\partial}{\partial x_i}$),

\begin{align}
& -3 x_7 \partial_{7} \bar{K}- x_9 \partial_{9}\bar{K} + x_5 \partial_{5} \bar{K} - x_6 \partial_{6}+3 x_4 \partial_{4} \bar{K} -2 x_2 \partial_{2} \bar{K} + x_8 \partial_{8} \bar{K} + 2 x_3 \partial_{3} \bar{K} =0 \\ \nonumber
&  - x_7 \partial_{6} \bar{K}  - 2 x_6 \partial_{5} \bar{K} - x_9 \partial_{8} \bar{K} -3 x_5 \partial_{4} \bar{K} =0 \\ \nonumber
& - x_4 \partial_{5} \bar{K} - x_8 \partial_{9} \bar{K} - 2 x_3 \partial_{1} \bar{K} + x_1 \partial_{2} \bar{K} - 2 x_5 \partial_{6} \bar{K} -3 x_6 \partial_{7} \bar{K} =0 \\ \nonumber
& -3 x_4 \partial_{1} \bar{K} +3 x_5 \partial_{2} \bar{K} =0 \\ \nonumber
& 2 x_6 \partial_{2} \bar{K} + x_4 \partial_{3} \bar{K} - x_5 \partial_{1} \bar{K} =0 \\ \nonumber
& x_6 \partial_{1} \bar{K} +2 x_5 \partial_{3} \bar{K} + x_7 \partial_{2} \bar{K} =0 \\ \nonumber
& 3 x_7 \partial_{1} \bar{K} +3 x_6 \partial_{3} \bar{K} =0 \\ \nonumber
& - x_8 \partial_{1} \bar{K} + x_9 \partial_{2} \bar{K} =0 \\ \nonumber
& x_9 \partial_{1} \bar{K} = x_8 \partial_{3} \bar{K} =0 
\end{align}

The last equation for $\{\bar{K},x_{10}\}$ is trivially satisfied. Using the Maple package to solve equations 4 and 8, we get
\[ \bar{K}=\bar{K}(x_3,...,x_{10}) \]
and then 5,6,7 and 9 give 
\[ \bar{K}=\bar{K}(x_4,...,x_{10}). \]
Next, 1,2 and 3 give that any function of the form $\bar{K}=\bar{K}(x_1,x_2,x_3,x_4)$

\begin{align}
&  \bar{K}_1= \frac{x_4 x_6-x_5^2}{x_4^{\frac{2}{3}}} , \label{k1b}\\
&   \bar{K}_2= \frac{x_4^2 x_7 -3 x_4 x_5 x_6 +2 x_5^3}{x_4} , \label{k2b}\\
&   \bar{K}_3= \frac{x_8}{x_4^{\frac{1}{3}}} ,\quad \label{k3b} \\
&\bar{K}_4= \frac{x_4 x_9 -x_5 x_8}{x_4^{\frac{2}{3}}}, \label{k4b} \\
&\bar{K}_5= x_{10} \label{k5b}
\end{align}

is a generalized invariant. We can associate the polynomial invariants $\tilde{K_i}$ in the following way

\begin{align}
& \tilde{K}_1= \bar{K}_5  \label{k1tb}\\
& \tilde{K}_2 = 4 \bar{K}_1^3 + \bar{K}_2^2 \label{k2tb}\\
&\tilde{K}_3 = - \bar{K}_1^2 \bar{K}_3^2 - \bar{K}_2 \bar{K}_3 \bar{K}_4 + \bar{K}_1 \bar{K}_4^2 \label{k3tb}\\
&\tilde{K}_4 =- \bar{K}_2 \bar{K}_3^3 + 3 \bar{K}_1 \bar{K}_3^2 \bar{K}_4 + \bar{K}_4^3 \label{k4tb}
\end{align}

Another type of generalized Casimir invariant $\hat{K}$ consists of a polynomial which commutes only in a given differential operator realization. The initial differential operators  with which we construct the commutation relations also provide extra information on the algebra, as an operator algebra. Applying the method 3, we get the following polynomials, ( where we use the realization as given by (\ref{real})):

\begin{align}
& \hat{K}_1= C_7  \label{k1a}\\
& \hat{K}_2= C_2^2- C_1 C_3 \label{k2a}\\
& \hat{K}_3=C_2C_3 - C_1 C_4 \label{k3a}\\
&\hat{K}_4=C_2 C_5 - C_1 C_6  \label{k4a}\\
& \hat{K}_5=C_3 C_5 - C_2 C_6  \label{k5a}\\
&\hat{K}_6= C_4 C_5 - C_3 C_6 \label{k5a}
\end{align}

Those relations indicate that, in this specific case, the further relations (\ref{k1a})-(\ref{k5a}) are valid in the realization. It was demonstrated how various realizations play a role for the Schr\"odinger and conformal Galilean algebra to get connection with special functions and PDEs relevant to physics in the context of fluid dynamics and holography. Using the initial realization (\ref{real}), we constructed the algebra providing only a vanishing expression for (\ref{k2}), (\ref{k3}) and (\ref{k4}). However, the realization from the vector field (\ref{vecf})  provides complicated PDEs.

\subsection{ Central extension of the subalgebra $V_2$ }

An approach to obtaining a centrally extended algebra consists in deforming an Abelian part of the radical of the Levi decomposition into a Heisenberg-type algebra by introducing a central element and preserving the Jacobi identity. We then consider 120 equations with 15 non-vanishing entries which then provide

\begin{equation}
 [C_1,C_2]=a_1, \quad [C_1,C_3]= a_2,\quad [C_2,C_3]=a_3
 \end{equation}
\[ [C_1,C_4]=a_4,\quad [C_2,C_4]=a_5,\quad [C_3,C_4]= a_6,\quad [C_5,C_6]=a_7 \]

and then the Jacobi identity gives us

\begin{equation}
 a_1=0,\quad a_2=0,\quad a_3=-\frac{a_4}{3},\quad a_5=0,\quad a_6=0 
\end{equation} 

Let us then use the notation $m_1=a_4$ and $m_2=a_7$. Applying method 4 and the NCalgebra package \cite{ncalg}, we find constraints on the parameters in the monomial expansion of a Casimir invariants of a given degree. We find the following 4th degree Casimir invariant, which now depends on both parts of the Levi decomposition, the semisimple part and the radical:
\begin{align}
 K&= -\frac{1}{2} m_1 m_2 J_0 + 5 m_2 C_1 C_4 - 2 m_2 C_2 C_3 + \frac{2}{3} m_1 C_5 C_6 - \frac{1}{6} m_1 m_2 J_0^2  \label{fK}  \\ \nonumber
&- \frac{2}{3} m_1 m_2 J_2 J_{-2} + m_2 J_0 C_1 C_4 -  m_2 J_0 C_2 C_3 + \frac{1}{3} m_1 J_0 C_5  C_6 + 2 m_2 J_2 C_1 C_3 -  2 m_2 J_2 C_2^2  \\ \nonumber
&+ \frac{1}{3} m_1 J_2 C_5^2 - 2 m_2 J_{-2} C_2 C_4 +  2 m_2 J_{-2} C_3^2 - \frac{1}{3} m_1 J_{-2} C_6^2 - \frac{3}{2 m_1} a_7 C_1^2 C_4^2 +    \frac{9}{m_1}  m_2 C_1 C_2 C_3 C_4  \\ \nonumber
&- \frac{6}{m_1}  m_2 C_1 C_3^3 +  C_1 C_3 C_6^2 - C_1 C_4  C_5 C_6 - \frac{6}{m_1} m_2 C_2^3 C_4 + \frac{9}{2 m_1}  m_2 C_2^2 C_3^2 -  C_2^2 C_6^2\\ \nonumber
& + C_2 C_3  C_5 C_6 + C_2 C_4 C_5^2 - C_3^2 C_5^2 \nonumber
\end{align}

We will now investigate and explain this Casimir invariant using a map called virtual copy. To our knowledge such cases, where the semisimple part $sl(2)$ acts on a combination of multiplets, is unexplored. In most known cases, such as for the Schr\"odinger and conformal Galilei algebras, the radical consists of only one multiplet. We can create the quadratic map in the enveloping algebra

\begin{equation}
J_0'= J_0 - \frac{3}{m_1} C_1 C_4 + \frac{3}{m_1} C_2 C_3 - \frac{1}{m_2} C_5 C_6 + \frac{5}{2} 
\end{equation}
\[ J_2'= J_2 + \frac{3}{m_1} C_2 C_4 - \frac{3}{m_1} C_3^2 + \frac{1}{2m_2} C_6^2 \]
\[ J_{-2}'= J_{-2} - \frac{3}{m_1} C_1 C_3 + \frac{3}{m_1} C_2^2 - \frac{1}{2m_2} C_5^2 \]

Then, $sl(2)$ is generated by $\{J_0',J_2', J_{-2}'\}$, where the commmutators with all $C_i$ vanish. Then the unique Casimir invariant is
\begin{equation}
 K'= J_0'^2 + 2 \{J_{2}',J_{-2}'\} 
\end{equation} 
and
\begin{equation}
 K= - \frac{m_1 m_2}{6} K' 
\end{equation}

\subsection{Realization and Casimir invariant}

The realizations, such as (\ref{vecf}), are usually not directly applicable in the case of central extensions. This is why we will rely on a similar construction as for the conformal Galilean algebra with central extensions. The realization for the centrally extended subalgebra takes the form

\begin{align}
&J_{-2}= t+ \frac{1}{2} m_2 t x_2^2 - \frac{1}{3} m_1 y_1^2 + x_1 \partial_{y_1} - t y_1 \partial_{y_1} - t x_2 \partial_{x_2} - 3 t x_1 \partial_{x_1} -t^2 \partial_t\\
&J_2= \partial_t\\
&J_0= 1 -y_1 \partial_{y_1} -x_2 \partial_{x_2} -3 x_1 \partial_{x_1} -2 t \partial_t\\
&C_1= \frac{1}{3} m_1 x_1 + m_1 t y_1 + 3 t^2 \partial_{y_1} -3 t^3 \partial_{x_1} \\
&C_2=\frac{1}{3} m_1 y_1 +2 t \partial_{y_1} -3 t^2 \partial_{x_1}\\
&C_3= \partial_{y_1} -3 t \partial_{x_1} \\
&C_4= -3 \partial_{x_1}\\
&C_5= -m_2 t x_2 + t \partial_{x_2}\\
&C_6= \partial_{x_2}\\
&C_7= \partial_t
\end{align}

Then, using this realization in the fourth-order Casimir invariant obtained via the virtual copy, we obtain the following

\begin{align}
&K=-\frac{1}{9}m_1(8 m_1 +3 m_2) - \frac{1}{3} ( m_1 -3 m_2)( t - m_2 x_2^2) \partial_{y_1}^2 + \frac{4}{9} m_1 (m_1-3 m_2) x_2 \partial_{x_2} \\ \nonumber
&-\frac{1}{18} m_1 (m_1 -3 m_2) x_2^2 \partial_{x_2}^2 - \frac{1}{3} m_1 (m_1 -3 m_2) (t -m_2 x_2^2) y_1 \partial_{x_1} + \frac{1}{9} m_1 ( m_1 -3 m_2) (t -m_2 x_2^2) \partial_t \nonumber
\end{align}

In the limit ($m_1 \rightarrow 0$ or $m_2 \rightarrow 0$), the realization is still valid. In the limit ($m_1 \rightarrow 0$ and $m_2 \rightarrow 0$) the Casimir vanishes identically.

\section{Other subalgebras based on the grading}

\subsection{Subalgebras $V_0$, $V_{-1}$ and $V_{12}$ }

In this subsection, we will consider the subalgebras $V_0$, $V_{-1}$ and $V_{12}$ and provide results obtained using the method 4, i.e. constructing Casimir from a monomial in the PBW basis and solving systems of linear equations. Those subalgebras and their radicals can also be put into multiplets of $sl(2)$, which requires a change of basis. The simple algebra $sl(2)$ acts as
\[  [sl(2),U_k] \subseteq U_k \]
Hence $U_k$, $k=-1,0,1,2$ are representations of $sl(2)$. Representations of $sl(2)$ are completely reducible. The explicit commutation relations are provided in Appendix B. For $V_{0}$, we have the following Casimir invariant

\begin{equation}
 K_1= S^0_1(0) - S^0_2(0) + S^0_3(0) 
\end{equation}

For the subalgebra $V_{-1}$ we have the Casimir invariant

\begin{align}
&K_1= S^{-1}(0) \\
& K_2= \frac{1}{4} T^{-1}(0)^2  + T^{-1}(2) T^{-1}(-2)\\
&K_3 = D^{-1}(1)^2 T^{-1}(-2) - D^{-1}(1)D^{-1}(-1)T^{-1}(0) + D^{-1}(-1)^2T^{-1}(2)
\end{align}

For the subalgebra $V_{12}$ we have the following Casimir

\begin{align}
&K_1= S^{2}(0)\\
&K_2=S^{1}(0) \\
&K_3= -T^1_1(0)^2 + T^1_1(2) T^1_1(-2) - Q^2(1)^2 T^1_1(-2) + Q^2(1) Q^2(-1) T^1_1(0)\\ \nonumber
&  + Q^2(1)Q^2(-3)T^1_1(2) + Q^2(3) Q^2(-1)T^1_1(-2)T^1_1(-2) - Q^2(3) Q^2(-3)T^1_1(0)\\ \nonumber
&- Q^2(-1)^2 T^1_1(2)\\ 
&  K_4= -Q^2(1)^2 T^1_2(-2) - Q^2(1) Q^2(-1)T^1_2(0) -Q^2(1) Q^2(-3) T^1_2(2) \\ \nonumber
&+ Q^2(3) Q^2(-1) T^1_2(-2) + Q^2(3)Q^2(-3) T^1_2(0) + Q^2(-1) Q^2(-1)T^1_2(2) \\
&K_5= -Q^2(1) D^2(1) T^1_1(-2)- 2 Q^2(1) D^2(-1) T^1_1(0) + Q^2(3) D^2(-1)T^1_1(-2) \\ \nonumber
& + 2 Q^2(-1) D^2(1)T^1_1(0) + Q^2(-1)D^2(-1)T^1_1(2) -Q^2(-3)D^2(1)T^1_1(2) \\ 
& K_6= D^2(1) D^2(1) T^1_1(-2) -2 D^2(1) D^2(-1) T^1_1(0) + D^2(-1) D^2(-1)T^1_1(2) \\ 
&K_7= 4 Q^2(1) Q^2(1) Q^2(1) Q^2(-3)- 3 Q^2(1) Q^2(1) Q^2(-1) Q^2(-1) \\ \nonumber
&- 6 Q^2(3) Q^2(1) Q^2(-1) Q^2(-3) + Q^2(3) Q^2(3) Q^2(-3) Q^2(-3)\\ \nonumber
& + 4 Q^2(3) Q^2(-1) Q^2(-1) Q^2(-1) \\
&K_8 = - Q^2(1) Q^2(1) D^2(-1) D^2(-1) + Q^2(1) Q^2(-1) D^2(1) D^2(-1)\\ \nonumber
&+ Q^2(1) Q^2(-3) D^2(1) D^2(1)+ Q^2(3) Q^2(-1) D^2(-1) D^2(-1)\\ \nonumber
& - Q^2(3) Q^2(-3) D^2(1) D^2(-1) - Q^2(-1) Q^2(-1) D^2(1) D^2(1)
\end{align}

\section{Other subalgebras with elements of different grading}

Here we will look at a different subalgebra, where there is an element of the different grading and also an interaction between those components.

\subsection{Subalgebra 1}

The subalgebra is spanned by $sl_2 \niplus R_1$, where the radical consists of elements of $U_0$, $U_{-1}$ and $U_1$

\[R_1= \{D^0_2(1),D^0_2(-1),S^{-1}(0), D^2(1), D^2(-1),D^1(1),D^1(-1) \} \]

with commutation relations

\begin{align}
&[J_2,J_0]=-2 J_2,\quad &[J_2,J_{-2}]=J_0, \quad &[J_0,J_{-2}]=-2 J_{-2} \nonumber \\
&[J_2,D^{0}_2(-1)]=D^{0}_2(1) ,\quad &[J_2,D^2(-1)]=D^2(1),\quad  &[J_2,D^1(-1)]=-D^1(1) \nonumber \\
&[J_0,D^0_2(1)]=D^0_2(1) ,\quad &[J_0,D^0_2(-1)]=- D^0_2(-1) ,\quad &[J_0,D^2(1)]=D^2(1) \nonumber \\
&[J_0,D^2(-1)]=-D^2(-1),\quad &[J_0,D^1(1)]=D^1(1),\quad &[J_0,D^1(-1)]=-D^1(-1) \nonumber \\
&[J_{-2},D^0_2(1)]=D^0_2(-1),\quad &[J_{-2},D^2(1)]=D^2(-1) ,\quad &[J_{-2},D^1(1)]=-D^1(-1) \nonumber 
\end{align}

Then the Casimir invariants are

\begin{align}
& K_1 = S^{-1}(0) \\
& K_2= D^0_2(1) D^2(-1) - D^{0}_2(-1) D^2(1) \\
& K_3= D^{0}_2(1) D^1(-1) + D^0_2(-1) D^1(1) \\
& K_4= D^2(1) D^1(-1) + D^2(-1) D^1(1) 
\end{align}

\subsection{Subalgebra 2}

We consider another example of a subalgebra which contains different components of the grading. This subalgebra takes the  form $sl_2 \niplus R_2$, where  $R_2=$ $\{$ $D^0_2(1)$, $D^0_2(-1)$, $T^{-1}(2)$, $T^{-1}(0)$ , $T^{-1}(-2)$, $Q^2(3)$, $Q^2(1)$, $Q^2(-1)$, $Q^2(-3)$, $D^1(1)$, $D^1(-1)$, $T^1_2(2)$, $T^1_2(0)$, $T^1_2(-2)$ $\}$.

The subalgebra has the following commutation relations

\begin{align}
&[J_2,J_0]=-2 J_2,\quad &[J_2,J_{-2}]=J_0, \quad &[J_0,J_{-2}]=-2 J_{-2} \nonumber \\
& [J_2,D^0_2(-1)]=D^0_2(1) ,\quad &[J_2,T^{-1}(0)]=-2 T^{-1}(2) ,\quad &[J_2,T^{-1}(-2)]=-T^{-1}(0) \nonumber \\
& [J_2,Q^2(1)]=Q^2(3),\quad &[J_2,Q^2(-1)]=2 Q^2(1) ,\quad &[J_2,Q^2(-3)]=3 Q^2(-1) \nonumber \\
& [J_2,D^1(-1)]=- D^1(1),\quad &[J_2,T^1_2(0)]=T^1_2(2) ,\quad &[J_2,T^1_2(-2)]=-2 T^1_2(0) \nonumber 
\end{align}

\begin{align}
& [J_0,D^0_2(1)]=D^0_2(1),\quad  &[J_0,D^0_2(-1)]= - D^0_2(-1),\quad &[J_0,T^{-1}(2)]=2 T^{-1}(2) \nonumber \\
& [J_0,T^{-1}(-2)]=-2 T^{-1}(-2),\quad &[J_0,Q^2(3)]=3 Q^2(3) ,\quad &[J_0,Q^2(1)]=Q^2(1) \nonumber \\
& [J_0,Q^2(-1)]=-Q^2(-1) ,\quad &[J_0,Q^2(-3)]=-3 Q^2(-3) ,\quad &[J_0,D^1(1)]=D^1(1) \nonumber \\
& [J_0,D^1(-1)]=- D^1(-1) ,\quad &[J_0,T^1_2(2)]=2 T^1_2(2) ,\quad &[J_0,T^1_2(-2)]=-2 T^1_2(-2) \nonumber \\
& [J_{-2},D^0_2(1)]=D^0_2(-1),\quad &[J_{-2},T^{-1}(2)]=- T^{-1}(1),\quad &[J_{-2},T^{-1}(0)]=-2 T^{-1}(-2) \nonumber \\
& [J_{-2},Q^2(3)]=3 Q^2(1),\quad &[J_{-2},Q^2(1)]=2 Q^2(-1),\quad &[J_{-2},Q^2(-1)]=Q^2(-3) \nonumber \\
& [J_{-2},D^1(1)]=-D^1(-1),\quad &[J_{-2},T^1_2(2)]=2 T^1_2(0),\quad &[J_{-2},T^1_2(0)]=-T^1_2(-2) \nonumber \\
& [J_{-2},T^1_2(0)]=-T^1_2(-2) ,\quad &[T^{-1}(2),D^1(-1)]=D^0_2(1), &[T^{-1}(0),D^1(1)]=-D^0_2(1) \nonumber \\ 
& [T^{-1}(0),D^1(-1)]=- D^0_2(-1) ,\quad &[T^{-1}(-2),D^1(1)]=D^0_2(-1) ,\quad &[D^1(1),T^1_2(-2)]=-Q^2(-1) \nonumber \\
& [D^1(-1),T^1_2(2)]=-Q^2(1),\quad &[D^1(-1), T^1_2(-2)]=Q^2(-3)  & \nonumber \\
\end{align}

We then have the Casimir invariant

\begin{align}
&K_1=  D^0_2(1) D^0_2(1) T^{-1}(-2) + D^0_2(1)  D^0_2(-1) T^{-1}(0) + D^0_2(-1) D^0_2(-1) T^{-1}(2) \\
&K_2= D^0_2(1) D^0_2(1) T^1_2(-2) + 2 D^0_2(1)  D^0_2(-1)  T^1_2(0) + D^0_2(1) T^{-1}(0) Q^2(-1)   \\
&+ D^0_2(1) T^{-1}(2) Q^2(-3) + D^0_2(1) T^{-1}(-2) Q^2(1) -  D^0_2(-1) D^0_2(-1) T^1_2(2)  \nonumber \\
&- D^0_2(-1) T^{-1}(0) Q^2(1) - D^0_2(-1) T^{-1}(2)  Q^2(-1) - D^0_2(-1) T^{-1}(-2) Q^2(3)  \nonumber \\
&K_3= D^0_2(1) Q^2(1) T^1_2(-2) + 2 D^0_2(1) Q^2(-1) T^1_2(0) - D^0_2(1) Q^2(-3) T^1_2(2)   \\
&- 2 D^0_2(-1) Q^2(1) T^1_2(0) -  D^0_2(-1) Q^2(3) T^1_2(-2) + D^0_2(-1) Q^2(-1) T^1_2(2)  \nonumber \\
&- T^{-1}(0) Q^2(1) Q^2(-1)  + T^{-1}(0) Q^2(3) Q^2(-3)+ 2 T^{-1}(2) Q^2(1) Q^2(-3) \nonumber \\
& - 2 T^{-1}(2) Q^2(-1)  Q^2(-1) - 2 T^{-1}(-2) Q^2(1) Q^2(1) + 2 T^{-1}(-2) Q^2(3) Q^2(-1)\nonumber 
\end{align}

\begin{align}
& K_4= -Q^2(1) Q^2(1) T^1_2(-2) - Q^2(1) Q^2(-1) T^1_2(0) - Q^2(1) Q^2(-3) T^1_2(2)  \\
& + Q^2(3) Q^2(-1) T^1_2(-2) + Q^2(3) Q^2(-3) T^1_2(0) + Q^2(-1) Q^2(-1) T^1_2(2) \nonumber \\
&K_5=D^0_2(1) D^0_2(1) D^0_2(1) Q^2(-3) - 3 D^0_2(1)  D^0_2(1) D^0_2(-1) Q^2(-1)  \\
&+ 3 D^0_2(1) D^0_2(-1) D^0_2(-1) Q^2_1 - D^0_2(-1) D^0_2(-1) D^0_2(-1) Q^2(3) \nonumber \\
&K_6= D^0_2(1) D^0_2(1) Q^2(1) Q^2(-3) - D^0_2(1) D^0_2(1) Q^2(-1) Q^2(-1)   \\
&+ D^0_2(1) D^0_2(-1) Q^2(1) Q^2(-1)- D^0_2(1) D^0_2(-1) Q^2(3) Q^2(-3)  \nonumber \\
& - D^0_2(-1) D^0_2(-1) Q^2(1) Q^2(1)+ D^0_2(-1)  D^0_2(-1) Q^2(3) Q^2(-1) \nonumber \\
&K_7=4 Q^2(1) Q^2(1) Q^2(1) Q^2(-3) - 3 Q^2(1)  Q^2(1) Q^2(-1) Q^2(-1)   \\
&- 6 Q^2(3) Q^2(1) Q^2(-1) Q^2(-3)+ Q^2(3) Q^2(3)  Q^2(-3) Q^2(-3)\nonumber \\
& + 4 Q^2(3) Q^2(-1) Q^2(-1) Q^2(-1) \nonumber
\end{align}

Using the realization given by (\ref{real}), the only non-vanishing Casimir invariants $K_1$ and $K_2$ are 

\[ K_1= x^2 \partial_5 \partial_6^2 f + x_1 x_2 \partial_1 \partial_6^2 f + x_1 \partial_3 \partial_6^2 \]

\[ K_2= 2 x_1 x_2^3 \partial_5 \partial_6^2 f -2 x_1^3 x_2 \partial_3 \partial_6^2 f -2 x_1 x_2^3 \partial_5 \partial_6^2 f + x_1^3 x_2 \partial_3 \partial_6 f \]

Those can be used to define PDEs and look into explicit representations of the algebras in terms of special functions. It is known that, for the Schr\"odinger and Conformal Galilean algebras, explicit realizations of the Casimir can lead to various generalizations of hypergeometric functions.

\section{ Structural properties of the Lie algebra W, Abelian subalgebras and labels}

Guided by the structure of multiplets, we will use another notation. The commutation relations take the form as given in Appendix B.  We introduce the nonation $M^j_k$

\[ M \in \{S,D,T,Q\} \]

Here, $k \in \{-1,0,1,2\}$, where k is the subscript of the corresponding subspace $U_k$ i.e. $M^i_k \subset U_k$ , j enumerates the multiplet of type M in $U_k$ if it has more than one of them. The members of the multiplet $M^j_k$ are distinguished by the value of m. It is the eigenvalue of the generator $J_0$, therefore

\[ [J_0,M^j_k(m)]=m M^j_k(m) \]

where 

\[ m= \begin{cases}
   0 & M=S \\
   \pm 1 & M=D \\
   \pm 2 , 0 & M=T \\
   \pm 3, \pm 1 & M=Q 
   \end {cases} \]
   
We then can write the following relations which represent the action of the generators $J_2$ and $J_{-2}$ which form the $sl(2)$ part of the Lie algebra  

\begin{equation}   
 [J_2,M^j_k(m)]=\lambda^j_k(m) M^j_k(m+2) 
\end{equation} 
\begin{equation} 
[J_{-2},M^j_k(m)]=\lambda^j_k(m) M^j_k(m-2) .
\end{equation}

The algebra is not perfect as the lower central series and upper central series never vanish

\[  [W,W]= W \setminus   \{S^1_0,S^2_0,S^3_0\} \]

\[  [W,[W,W]]= W \setminus   \{S^1_0,S^2_0,S^3_0\} \]

The maximal set of commuting singlets is given by 

\[ S^{-1}_1,\quad S^0_1,\quad S^0_2 \]

We now extend the $sl(2)$ and related Cartan associated eigenvalues with three singlets  and use their value to label components of every multiplets where $ X \in W$

\begin{equation}
 [J_0,X]= m  X ,\quad [S^{-1},X]= a X ,\quad [S^0_1,X]= b X ,\quad [S^0_2,X]= c X 
\end{equation}

By including the label i, which is the grading label we get

\[ (i,m,a,b,c) \]

We then label a given element and obtain the property

\begin{equation}
 [(i,m,a,b,c),(i',m',a',b',c')]= [(i+i'),(m+m'),(a+a'),(b+b'),c+c')] 
\end{equation} 

\par
%
%
\section{Conclusion}

In this paper we have presented the differential operator realization and explicit commutation relations for the Lie algebra of the lowest transitively differential group of degree three in two different bases, which are described in Appendices A and B.  We have presented decompositions, different from the Levi decomposition, which has allowed us to provide an alternative basis in terms of multiplets of the semisimple part $sl(2)$. The commutation relations of Appendix B explicitly refer to the Lie structure in terms of multiplets with respect to the semisimple part. \par

We have also presented a grading which is unbalanced involving $U_{-1}$, $U_{0}$, $U_1$ and $U_{2}$, which then allows us to present a different subalgebra of interest. This leads to a different algebra, where the radical consists of a reducible representation of $sl(2)$, which is already of interest with regard to the theory of Casimir invariants. 

We have presented 5 different methods for constructing Casimir invariants ( and generalized Casimir) and applied those methods to the subalgebra $V_2$. We have also considered a central extension and constructed a fourth-degree Casimir invariant, derived via the method of virtual copy of Lie algebras. We also present Casimir invariants for the other three subalgebras $V_0$, $V_{12}$ and $V_{-1}$. 

Beyond the grading we provided a suitable set of commuting generators, which can then provide further insight into the decomposition of the algebra.

%
%
.\par
%
%
\section*{Acknowledgments}

We would like to dedicate this paper to the memory of our mentor Jiri Patera (Universit\'e de Montr\'eal) who passed away in January 2022. He was always a source of inspiration and encouragement.

AMG was supported by a research grant from NSERC of Canada. IM was supported by the Australian Research Council Fellowship FT180100099. I.M. thanks the Centre de Recherches Math\'ematiques, Universit\'e de Montr\'eal for its hospitality.  \par
%
%

\section{Appendix A}

\begin{align}
&[A_1,L_1]= A_6,\quad &[A_1,L_7]= A_3,\quad &[A_1,L_9]=A_4 ,\quad &[A_1,L_9]= A_4, \nonumber \\ \nonumber
&[A_1,L_{11}]=A_5,\quad &[A_1,L_{13}]=A_1,\quad &[A_1,L_{15}]= A_2,\quad &[A_1,B_1]=2 L_1, \\ \nonumber
&[A_1,B_2]= L_2 ,\quad &[A_1,B_4]= L_5,\quad & [A_1,B_5]= - L_4,\quad &[A_1,B_6]=\frac{1}{2}L_{10}, \\ \nonumber
&[A_1,B_7]= 2 L_{12}+ 2 L_9 ,\quad &[A_1,B_8]= 2 L_8 ,\quad &[A_1,B_9]= \frac{1}{2} L_{10} +2 L_7,\quad &[A_1,C_1]= 3 B_1,\\ \nonumber
&[A_1,C_2]=2 B_2,\quad &[A_1,C_3]= B_3,\quad &[A_1,C_5]= B_6 + 3 B_9 ,\quad &[A_1,C_6]= B_8 \\ \nonumber
&[A_1,C_7]=2 B_4 ,\quad &[A_2,L_2]=A_6,\quad &[A_2,L_8]= A_3,\quad &[A_2,L_{10}]= A_4,\\ \nonumber
&[A_2,L_{12}]= A_5,\quad &[A_2,L_{14}]=A_1,\quad &[A_2,L_{16}]= A_2,\quad & [A_2,B_2]= L_1,\\ \nonumber
&[A_2,B_3]=2 L_2,\quad &[A_2,B_3]= 2 L_2 ,\quad &[A_2,B_4]= -L_4,\quad &[A_2,B_5]=L_3,\\ \nonumber
&[A_2,B_6]=\frac{1}{2} L_9 +2 L_{12},\quad &[A_2,B_7]= 2 L_{11} ,\quad &[A_2,B_8]= 2 L_{10}+ 2 L_7 ,\quad &[A_2,B_9]= \frac{1}{2} L_9, \\ \nonumber
&[A_2,C_2]= B_1,,\quad &[A_2,C_3]= 2 B_2,\quad &[A_2,C_4]= 3 B_3,\quad  &[A_2,C_5]= B_7 , \\ \nonumber
&[A_2,C_6]= B_9 +3 B_6 ,\quad &[A_2,C_7]= 2 B_5 ,\quad  &[A_3,L_3]= A_6,\quad &[A_3,L_{13}]= 2 A_3 ,\\ \nonumber
 &[A_3,L_{15}]= A_4,\quad &[A_3,A_{17}]=A_3 ,\quad &[A_3,B_5]=L_2 ,\quad &[A_3,C_7]=B_3 ,\\ \nonumber
&[A_4,L_4]= A_6,\quad &[A_4,L_{13}]=A_4,\quad &[A_4,L_{14}]= 2 A_3,\quad &[A_4,L_{15}]= 2 A_5,\\ \nonumber
&[A_4,L_{16}]= A_4,\quad &[A_4,L_{17}]= A_4 ,\quad &[A_4,B_4]=-L_2,\quad &[A_4,B_5]=-L_1 , \\ \nonumber
&[A_4,C_7]=-2 B_2,\quad &[A_5,L_5]=A_6,\quad &[A_5,L_{14}=A_4,\quad &[A_5,L_{16}]=2 A_5 ,\\ \nonumber
&[A_5,L_{17}]= A_5,\quad &[A_5,B_4]=L_1,\quad &[A_5,C_7]=B_1,\quad &[A_6,L_6]=A_6,\\ \nonumber
&[L_1,L_6]=L_1,\quad &[L_1,L_{13}]=-L_1 ,\quad &[L_1,L_{14}]=-L_2,\quad &[L_2,L_6]= L_2,\\ \nonumber
&[L_2,L_{15}]=-L_1,\quad &[L_2,L_{16}]= -L2,\quad &[L_3,L_7]= -L1,\quad &[L_3,L_8]=-L2 ,\\ \nonumber
&[L_3,L_{13}]=-2L_3,\quad &[L_3,L_{14}]=-2L_4, &[L_3,L_{17}]=-L_3,\quad  &[L_3,B_8]= -2 B_2,\\ \nonumber
&[L_3,B_9]= - B_1 ,\quad &[L_3,C_5=-C_1 ,\quad &[L_3,C_6]= -C_2 ,\quad &[L_4,L_6]=L_4 , \\ \nonumber
&[L_4,L_9]=-L_1,\quad &[L_4,L_{10}]=-L_2,\quad &[L_4,L_{13}]=-L_4,\quad &[L_4,L_{14}]= -L_5,\\ \nonumber
&[L_4,L_{15}]=-L_3 ,\quad &[L_4,L_{16}]=-L_4,\quad &[L_4,L_{17}]=-L_4 ,\quad  &[L_{4},B_6]=-\frac{1}{2} B_2,\\ \nonumber
&[L_4,B_7]= - B_1 ,\quad &[L_4,B_8]=-B_3,\quad &[L_4,B_9]=-\frac{1}{2} B_2,\quad  &[L_4,C_5]=-C_2,\\ \nonumber
&[L_4,C_6]= -C_3 ,\quad &[L_5,L_{11}]= -L_1,\quad &[L_5,L_{12}]=-L_2,\quad &[L_5,L_{15}]= -2 L_4,
\end{align}

\begin{align}
&[L_5,L_{16}]=-2 L_5  ,\quad &[L_5,L_{17}]=-L5,\quad &[L_5,B_6 ]=-B_3,\quad &[L_5,B_7]-2 B_2, \nonumber \\ \nonumber
&[L_5,C_5]= -C_3,\quad &[L_5,C_6=-C_4 ,\quad &[L_6,C_1]=-C_1,\quad &[L_6,C_2]=-C_2,\\ \nonumber
&[L_6,C_3]=-C_3,\quad  &[L_6,C_4]= -C_4,\quad &[L_6,C_7]-C_7 ,\quad &[L_7,L_{13}]=L_7,\\ \nonumber
&[L_7,L_{14}]=- L_8 ,\quad &[L_7,L_{15}]=L_9 ,\quad &[L_7,L_{17}]=L_7 ,\quad  &[L_7,B_5]=B_2,\\ \nonumber
&[L_7,C_7]=C_3,\quad &[L_8,L_{13}]=2 L_8,\quad &[L_8,L_{15}]=L_{10} -L_7 ,\quad &[L_8,L_{16}]=- L_8,\\ \nonumber
&[L_8,L_{17}]=L_8 ,\quad  &[L_8,B_5]= B_3,\quad &[L_8,C_7]=C_4,\quad &[L_9,L_{14}]-L_{10}+ 2 L_{7},\\ \nonumber
&[L_9,L_{15}]=2 L_{11},\quad &[L_9,L_{16}]=L_9,\quad &[L_{9},L_{17}]=L_9 ,\quad  &[L_9,B_4]=-B_2,\\ \nonumber
&[L_9,B_5]=-B_1,\quad &[L_9,C_7]=-2 C_2,\quad &[L_{10},L_{13}]=L_{10},\quad &[L_{10},L_{14}]=2 L_8,\\ \nonumber
&[L_{10},L_{15}]= -L_9+ 2 L_{12} ,\quad &[L_{10},L_{17}]=L_{10},\quad &[L_{10},B_4]=-B_3,\quad &[L_{10},B_5]=-B_2,\\ \nonumber
& [L_{10},C_7]=-2 C_3,\quad &[L_{11},L_{13}]= - L_{11},\quad &[L_{11},L_{14}]=L_9 - L_{12},\quad &[L_{11},L_{16}]=2 L_{11},\\ \nonumber
&[L_{11},L_{17}]= L_{11} ,\quad &[L_{11},B_4]=B_1 ,\quad &[L_{11},C_7]= C_1,\quad &[L_{12},L_{15}]=- L_{11},\\ \nonumber
&[L_{12},L_{16}]= L_{12} ,\quad &[L_{12},L_{17}]= L_{12},\quad &[L_{12},B_4]=B_2,\quad &[L_{12},C_7]=C_2,\\ \nonumber
&[L_{13},L_{14}]=-L_{14},\quad &[L_{13},L_{15}]=L_{15},\quad &[L_{13},B_1]= 2 B_1,\quad  &[L_{13},B_2]= B_2,\\ \nonumber
& [L_{13},B_4]= B_4,\quad &[L_{13},B_5]=2 B_5,\quad  &[L_{13},B_8]= - B_8,\quad &[L_{13},C_1]= 3 C_1,\\ \nonumber
&[L_{13},C_2]=2 C_2,\quad &[L_{13},C_5]= C_5,\quad &[L_{13},C_7]=2 C_7 ,\quad &[L_{14},L_{15}]=L_{16}-L_{13} ,\\ \nonumber
&[L_{14},L_{16}]=-L_{14},\quad &[L_{14},B_1]=2 B_2, \quad &[L_{14},B_2]=B_3,\quad &[L_{14},B_5]= - B_4 ,\\ \nonumber
&[L_{14},B_6]= -\frac{1}{2} B_8,\quad &[L_{14},B_7]=-2 B_9 +2 B_6,\quad &[L_{14},B_9]= \frac{1}{2} B_8,\quad &[ L_{14},C_1]= 3 C_2 ,\\ \nonumber
&[L_{14},C_2]=2 C_3,\quad &[L_{14},C_3]= C_4,\quad &[L_{14},C_5]= C_6 ,\quad &[L_{15},L_{16}]=L_{15},\\ \nonumber
&[L_{15},B_2]= B_1 ,\quad &[L_{15},B_3]= 2 B_2,\quad &[L_{15},B_4]=-B_5,\quad &[L_{15},B-6]=\frac{1}{2} B_7 ,\\ \nonumber
&[ L_{15},B_8]= -2 B_6 + 2 B_9 ,\quad &[L_{15},B_9]= -\frac{1}{2}B_7 ,\quad &[L_{15},C_2]=C_1,\quad &[L_{15},C_2]= C_1,\\ \nonumber
&[L_{15},C_3]= 2 C_2,\quad &[L_{15},C_4]=3 C_3,\quad  &[L_{15},C_6]=C_5,\quad &[L_{16},B_2]=B_2,\\ \nonumber
&[L_{16},B_3]= 2 B_3,\quad &[L_{16},B_4]= 2 B_4,\quad &[L_{16},B_7]-B_7] ,\quad &[L_{16},B_8]=B_8,\\ \nonumber
&[L_{16},C_2]=C_2,\quad &[L_{16},C_3]=2 C_3,\quad &[L_{16},C_4]=3 C_4 ,\quad &[L_{16},C_6]=C_6,\\ \nonumber
&[L_{16},C_7]= 2 C_7 ,\quad &[L_{17},B_4]= B_4,\quad  &[L_{17},B_5]=B_5,\quad &[L_{17},B_6]=-B_6 ,\\ \nonumber
&[L_{17},B_7]= - B_7 ,\quad &[L_{17},B_8]= -B_8 ,\quad &[L_{17},B_9]= - B_9 ,\quad &[L_{17},C_5]=-C_5,\\ \nonumber
&[L_{17},C_6]=-C_6,\quad &[L_{17},C_7]=C_7,\quad &[B_4,B_6]=-\frac{1}{2} C_3 ,\quad &[B_4,B_7]=-C_2,\\ \nonumber
\end{align}

\begin{align}
&[B_4,B_8]= C_4,\quad &[B_4,B_9]=\frac{1}{2} C_3, \quad  &[B_5,B_6]= \frac{1}{2} C_2,\quad &[B_5,B_6]= C_1,\\ \nonumber
&[B_5,B_7]= C_1,\quad &[B_5,B_8]= - C_3,\quad &[B_5,B_9]= - \frac{1}{2} C_2. \nonumber
\end{align}

\section{Appendix B}

\begin{align}
&[J_2,J_0]=-2 J_2,\quad &[J_2,J_{-2}]=J_0,\quad  &[J_2,T^{0}(-2)]= 2 T^0(0),\nonumber \\ \nonumber
&[J_2,T^0(0)]=T^0(2),\quad &[J_2,D^0_2(-1)]=D^0_2(1) ,\quad &[J_2,D^{-1}(-1)]=- D^{-1}(1),\\ \nonumber
&[J_2, T^{-1}(0)]=-2 T^{-1}(2) , \quad &[J_2,T^{-1}(-2)]=- T^{-1}(0),\quad &[J_2,Q^{2}(1)]=Q^{2}(3)  \\ \nonumber
&[J_2,Q^{2}(-1)]=2 Q^{2}(1) ,\quad &[ J_2,Q^{2}(-3)]= 3 Q^2(-1) ,\quad &[J_2,D^2(-1)]=D^2(1) \\ \nonumber
& [J_2,T^1_1(0)]=2 T^1_1(2),\quad &[J_2,T^1_1(-2)]=2 T^1_1(0) ,\quad &[J_2,D^1(-1)]=- D^1(1)\\ \nonumber
&[J_2,T^1_2(0)]=-2 T^1_2(2),\quad &[J_2,T^1_2(-2)]=-2 T^1_2(0) ,\quad &[J_2,Q^0(3)]=Q^0(1)\\ \nonumber
&[J_2,Q^0(1)]=Q^0(-1), \quad &[J_2,Q^0(-1)]=Q^0(-3) ,\quad &[J_2,D^0_1(1)]=-D^0_1(-1)\\ \nonumber
&[J_2,Q^0(1)]=Q^0(-1), \quad &[J_2,Q^0(-1)]=Q^0(-3) ,\quad &[J_2,D^0_1(1)]=-D^0_1(-1)\\ \nonumber
&[J_0,J_{-2}]=-2 J_{-2} ,\quad &[J_0,T^0(2)]=2 T^0(2) ,\quad &[J_0,T^0(-2)]= -2 T^0(-2)\\ \nonumber
&[J_0,D^0_2(1)]=D^0_2(1) ,\quad &[J_0,D^0_2(-1)]=-D^0_2(-1) ,\quad &[J_0,D^{-1}(1)]=D^{-1}(1)\\ \nonumber
&[J_0,D^{-1}(-1)]=D^{-1}(-1),\quad &[J_0,T^{-1}(2)]=2 T^{-1}(2) ,\quad &[J_0,T^{-1}(-2)]=-2 T^{-1}(-2)\\ \nonumber
&[ J_0,Q^2(3)]=3 Q^2(3),\quad &[J_0,Q^2(1)]= Q^2(1) ,\quad &[J_0,Q^2(-1)]=- Q^2(-1)\\ \nonumber
&[J_0,Q^2(-3)]=(-3) Q^2(-3) ,\quad &[J_0,D^2(1)]=D^2(1),\quad &[J_0,D^2(-1)]=- D^2(-1)\\ \nonumber
&[J_0,T^1_1(2)]=2 T^1_1(2) ,\quad &[J_0,T^1_1(-2)]=-2 T^1_1(-2),\quad &[J_0,D^1(1)]= D^1(1)\\ \nonumber
&[J_0,D^1(-1)]=- D^1(-1) ,\quad &[J_0,T^1_2(2)]= 2 T^1_2(2) ,\quad &[J_0,T^1_2(-2)]=-2 T^1_2(-2)\\ \nonumber
&[J_0,Q^0(3)]=-3 Q^0(3) ,\quad &[J_0,Q^0(1)]=- Q^0(1) ,\quad &[J_0,Q^0(-1)]=Q^0(-1)\\ \nonumber
&[J_0,Q^0(-3)]= 3 Q^0(-3),\quad &[J_0,D^0_1(1)]=- D^0_1(1) ,\quad &[J_0,D^0_1(-1)]=D^0_1(-1) \\ \nonumber
& [J_{-2},T^0(0)]= T^0(-2),\quad &[J_{-2},T^0(2)]=2 T^0(0),\quad &[J_{-2},T^0(-2)]= T^0(-2) \\ \nonumber
&[J_{-2},D^0_2(1)]=D^0_2(-1),\quad &[J_{-2},D^{-1}(1)]=-D^{-1}(-1),\quad &[J_{-2},T^{-1}(2)]=- T^{-1}(0)\\ \nonumber
&[J_{-2},T^{-1}(0)]=-2 T^{-1}(-2),\quad &[J_{-2},Q^{2}(3)]=3 Q^2(1),\quad &[J_{-2},Q^2(1)]= 2 Q^2(-1),\\ \nonumber
&[J_{-2},Q^2(-1)]=Q^2(-3),\quad &[J_{-2},D^2(1)]= D^2(-1),\quad &[J_{-2},T^1_1(2)]=2 T^1_1(0),\\ \nonumber
&[ J_{-2},T^1_1(0)]= T^1_1(-2),\quad &[J_{-2},D^1(1)]=- D^1(-1),\quad &[J_{-2},T^1_2(2)]=- T^1_2(-2),\\ \nonumber
&[J_{-2},Q^0(1)]=3 Q^0(3),\quad &[J_{-2},Q^0(-1)]=4 Q^0(1),\quad &[J_{-2},Q^0(-3)]=3 Q^0(-1),\\ \nonumber
& [J_{-2},D^0_1(-1)]=- D^0_1(1),\quad &[T^0(0),S^0_3(0)]=T^0(0),\quad &[T^0(0),T^{-1}(0)]=- S^{-1}(0) ,\\ \nonumber
&[T^0(0),D^2(1)]=- Q^2(1),\quad &[T^0(0),D^2(-1)]=- Q^2(-1),\quad &[T^0(0),S^1(0)]- T^1_1(0), \nonumber
\end{align}

\begin{align}
&[T^0(0),T^1_2(2)]=-T^1_1(2),\quad &[T^0(2),T^0(-2)]=0,\quad &[T^0(0),T^1_2(-2)]= T^1_1(-2),\nonumber \\ \nonumber
& [T^0(0),Q^0(1)]=D^0_2(-1),\quad &[T^0(0),Q^0(-1)]=2 D^{0}_2(1),\quad &[T^0(0),D^0_1(1)]= - D^0_2(-1) ,\\ \nonumber
&[T^0(0),D^0_1(-1)]=- D^0_2(1),\quad &[T^0(0),S^0_1(0)]=-2 T^0(0),\quad &[T^0(0),S^0_2(0)]=- T^0(0) ,\\ \nonumber
&[T^0(2),S^0_3(0)]=T^0(2) ,\quad &[T^0_2,T^{-1}(-2)]=- S^{-1}(0),\quad &[T^0(2),D^2(1)]=- Q^2(3) ,\\ \nonumber
&[T^0(2),D^2(-1)]=- Q^2(1) ,\quad &[T^0(2),S^1(0)]=-T^1_1(2),\quad &[T^0(2),T^0(-2)]=0,\\ \nonumber
& [T^0(2),T^1_2(-2)]=-2 T^1_1(0) ,\quad &[T^0(2),Q^0(3)]=- D^0_2(-1) ,\quad &[T^0(2),Q^0(1)]=- D^0_2(1)\\ \nonumber
&[T^0(2),D^0_1(1)]=-2 D^0_2(1),\quad &[T^0(2),S^0_1(0)]=-2 T^0(2) ,\quad &[T^0(2),S^0_2(0)]=- T^0(2)\\ \nonumber
&[T^0(-2),S^0_3(0)]=T^0(-2),\quad &[T^0(-2),T^{-1}(2)]=-S^{-1}(0),\quad &[T^0(-2),D^2(1)]=-Q^2(-1)\\ \nonumber
& [T^0(-2),D^2(-1)]=-Q^2(-3),\quad &[T^0(-2),S^1(0)]=-T^1_1(-2),\quad &[T^0(-2),T^1_2(2)]=-2 T^1_1(0)\\ \nonumber
&[T^0(-2),T^1_2(0)]=-T^1_1(-2),\quad &[T^0(-2),Q^0(-1)]=-2 D^0_2(-1),\quad &[T^0(-2),Q^0(-3)]=-6 D^0_2(1)\\ \nonumber
&[T^0(-2),D^0_1(-1)]=-2D^0_2(-1),\quad &[T^0(-2),S^0_1(0)]=-2 T^0(-2),\quad &[ T^0(-2),S^0_2(0)]= - T^0(-2)\\ \nonumber
&[D^0_2(1),S^0_3(0)]= D^0_2(1) ,\quad &[D^0_2(1),D^{-1}(-1)]=- S^{-1}(0),\quad
&[D^0_2(1),S^0_1(0)]=- D^0_2(1)\\ \nonumber
&[D^0_2(-1),S^0_3(0)]= D^0_2(-1) ,\quad &[D^0_2(-1),D^{-1}(1)]=- S^{-1}(0) ,\quad  
&[ D^0_2(-1),S^0_1(0)]=- D^0_2(-1)\\ \nonumber
&[S^0_3(0),S^{-1}(0)]=- S^{-1}(0) ,\quad 
&[S^0_3(0),Q^2(3)]=- Q^2(3),\quad &[S^0_3(0),Q^2(1)]=- Q^2(1)\\ \nonumber
& [S^0_3(0),Q^2(-1)]=- Q^2(-1),\quad &[S^0_3(0),Q^2(-3)]=- Q^2(-3),\quad &[S^0_3(0),S^2(0)]=- S^2(0)\\ \nonumber
&[S^0_3(0), T^1_1(2)]=- T^1_1(2),\quad &[S^0_3(0),T^1_1(0)]=-T^1_1(0),\quad &[S^0_3(0),T^1_1(-2)]= - T^1_1(-2)\\ \nonumber
&[S^0_3(0),D^1(1)]=-D^1(1),\quad &[S^0_3(0),D^1(-1)]=-D^1(-1) ,\quad
&[D^{-1}(1) ,Q^2(1)]=T^1_1(2) \\ \nonumber
&[D^{-1}(1),Q^2(-1)]= 2 T^1_1(0) ,\quad &[D^{-1}(1),Q^2(-3)]=3 T^1_1(-2),\quad &[ D^{-1}(1),D^2(1)]=T^1_2(2) \\ \nonumber
&[D^{-1}(1),D^2(-1)]= T^1_2(0) + 2 S^1(0) ,\quad &[D^{-1}(1),S^2(0)]= 2 D^1(1) ,\quad &[D^{-1}(1),T^1_1(0)]=2 D^0_2(-1)\\ \nonumber
&[D^{-1}(1),D^1(1)]=T^0(2),\quad &[D^{-1}(1),D^1(-1)]=- T^0(0) ,\quad &[D^{-1}(1),S^1(0)]=D^0_1(-1)\\ \nonumber
&[ D^{-1}(1),Q^0(3)]= T^{-1}(-2) ,\quad &[D^{-1}(1),Q^0(1)]=- T^{-1}(0)],\quad 
&[D^{-1}(1),Q^0(-1)]=2 T^{-1}(2)\\ \nonumber
&[ D^{-1}(1),D^0_1(1)]= T^{-1}(0),\quad &[ D^{-1}(1),D^0_1(-1)]=2 T^{-1}(2) ,\quad &[ D^{-1}(1),S^0_1(0)]= D^{-1}(1)\\ \nonumber
&[D^{-1}(-1),Q^2(3)]= 3 T^1_1(2),\quad &[D^{-1}(-1),Q^2(1)]=2 T^1_1(0),\quad &[D^{-1}(-1),Q^2(-1)]=T^1_1(-2)\\ \nonumber
&[D^{-1}(-1),D^2(1)]=- T^1_2(0) +2 S^1(0) ,\quad &[D^{-1}(-1),D^2(-1)]=T^1_2(-2),\quad &[D^{-1}(-1),S^2(0)]=2 D^{-1}(-1)\\ \nonumber
&[D^{-1}(1),T^1_2(-2)]= -\frac{1}{3}( Q^0(1)-4 D^0_1(1)) &    & \nonumber
\end{align}

\begin{align}
&[D^{-1}(-1),T^1_1(2)]=2 D^0_2(1),\quad &[D^{-1}(-1),T^1_1(0)]=D^0_2(-1),\quad &[D^{-1}(-1),D^1(1)]=- T^0(0) \nonumber \\ \nonumber
&[D^{-1}(-1),Q^0(-1)]=-2 T^{-1}(0),\quad &[D^{-1}(-1),Q^0(-3)]=6 T^{-1}(2),\quad &[D^{-1}(-1),D^0_1(1)]=2 T^{-1}(-2)\\ \nonumber
&[D^{-1}(-1),D^0_1(-1)]=T^{-1}(0),\quad &[D^{-1}(-1),S^0_1(0)]=D^{-1}(-1) ,\quad &[T^{-1}(2),S^2(0)]=T^1_1(2)\\ \nonumber
&[D^{-1}(-1),D^0_1(-1)]=T^{-1}(0),\quad &[D^{-1}(-1),S^0_1(0)]=D^{-1}(-1) ,\quad &[T^{-1}(2),S^2(0)]=T^1_1(2)\\ \nonumber
&[T^{-1}(2),D^1(-1)]=D^0_2(1),\quad &[T^{-1}(2),S^0_1(0)]=2 T^{-1}(2),\quad &[T^{-1}(2),S^0_2(0)]=T^{-1}(2)\\ \nonumber
&[T^{-1}(0),S^2(0)]=-2 T^1_1(0),\quad &[T^{-1}(0).D^1(1)]=- D^0_2(1),\quad &[T^{-1}(0),D^{1}(-1)]= - D^0_2(-1)\\ \nonumber
&[T^{-1}(0),S^0_1(0)]=2 T^{-1}(0),\quad &[T^{-1}(0),S^0_2(0)]= T^{-1}(0),\quad &[T^{-1}(-2),S^{2}(0)]=T^1_1(-2)\\ \nonumber
&[T^{-1}(-2),D^1(1)]=D^0_2(-1),\quad &[T^{-1}(-2),S^0_1(0)]=2 T^{-1}(-2),\quad &[T^{-1}(-2),S^0_2(0)]= T^{-1}(-2)\\ \nonumber
&[Q^2(3), D^0_1(-1)]= -3 Q^2(3),\quad &[Q^{2}(1),S^0_1(0)]=-3 Q^2(1),\quad &[Q^2(-1),S^0_1(0)]=-3 Q^2(-1)\\ \nonumber
&[Q^2(-3),S^0_1(0)]=- Q^2(-3),\quad &[D^2(1),S^0_2(0)]=D^2(1),\quad &[D^2(1),S^0_2(0)]= D^2(1)\\ \nonumber
&[D^2(-1),S^0_1(0)]=- D^2(-1),\quad &[D^2(-1),S^0_2(0)]=D^2(-1),\quad  &[S^2(0),Q^0(3)]=-Q^2(-3)\\ \nonumber
& [S^2(0),Q^0(1)]=-3 Q^2(-1),\quad &[S^2(0),Q^0(-1)]=-6 Q^2(1),\quad &[S^2(0),Q^0(-3)]=-6 Q^2(3)\\ \nonumber
& [S^2(0),S^0_1(0)]=-4 S^2(0),\quad &[S^2(0),S^0_2(0)]=- S^2(0),\quad &[T^1_1(2),S^0_1(0)]=-2 T^1_1(2)\\ \nonumber
&[T^1_1(0),S^0_1(0)]=-2 T^1_1(0),\quad 
&[T^1_1(-2),S^0_1(0)]=-2 T^1_1(-2),\quad &[D^1(1),T^1_2(2)]=Q^2(3)\\ \nonumber
&[D^1(1),T^1_2(-2)]=- Q^2(-1),\quad &[D^1(1),Q^0(3)]=-T^1_1(-2),\quad &[D^1(1),Q^0(1)]=-2 T^1_1(0)\\ \nonumber
&[D^1(1),Q^0(-1)]=-2 T^1_1(2),\quad &[D^1(1),D^0_1(1)]=- T^1_1(0),\quad &[D^1(1),D^0_1(-1)]=T^1_1(2)\\ \nonumber
&[D^1(-1),T^1_2(2)]=-Q^2(1),\quad &[D^1(-1),T^1_2(0)]=-Q^2(-1),\quad &[D^1(-1),T^1_2(-2)]= Q^2(-3)\\ \nonumber
&[D^1(-1),Q^0(1)]=- T^1_1(-2),\quad &[D^1(-1),Q^0(-3)]=-4 T^1_1(0),\quad &[D^1(-1),Q^0(-3)]=-6 T^1_1(2)\\ \nonumber
&[D^1(-1),D^0_1(1)]= T^1_1(-2),\quad &[ D^1(-1),D^0_1(-1)]=- T^1_1(0),\quad &[D^1(-1),S^0_1(0)]=-3 D^1(-1)\\ \nonumber
& [D^1(-1),S^0_2(0)]=- D^1(-1),\quad &[S^1(0),S^0_2(0)]=S^1(0),\quad &[T^1_2(2),S^0_2(0)]=T^1_2(2)\\ \nonumber
&[D^1(-1),S^0_2(0)]=- D^1(-1),\quad &[S^1(0),S^0_2(0)]=S^1(0),\quad &[T^1_2(2),S^0_2(0)]=T^1_2(2)\\ \nonumber
&[T^1_2(0),S^0_2(0)]=T^1_2(0),\quad &[T^1_2(-2),S^0_2(0)]=T^1_2(-2),\quad &[Q^0(3),S^0_1(0)]=Q^0(3)\\ \nonumber
&[Q^0(3),S^0_2(0)]=Q^0(3),\quad &[Q^0(1),S^0_1(0)]=Q^0(1),\quad &[Q^0(1),S^0_2(0)]=Q^0(1)\\ \nonumber
& [Q^0(-1),S^0_1(0)]=Q^0(-1),\quad &[Q^0(-1),S^0_2(0)]=Q^0(-1),\quad &[Q^0(-3),S^0_1(0)]=Q^0(-3) \nonumber
\end{align}

\begin{align}
& [Q^0(-3),S^0_2(0)]=Q^0(-3),\quad &[D^0_1(1),S^0_1(0)]=D^0_1(1),\quad &[D^0_1(1),S^0_2(0)]= D^0_1(1) \nonumber \\ \nonumber
&[D^0_1(1),S^0_2(0)]=D^0_1(1),\quad &[D^0_1(-1),S^0_1,(0)]=D^0_1(-1),\quad &[D^0_1(-1),S^0_2(0)]=D^0_1(-1)\\ \nonumber
&[D^{-1}(-1),T^1_2(0)]=-\frac{2}{3}D^0_1(1)-\frac{2}{3} Q^0(1),\quad &[D^{-1}(-1),T^1_2(-2)]=2 Q^0(3),\\ \nonumber
&[D^{-1}(-1),Q^0(1)]=T^{-1}(-2) \\ \nonumber
&[D^{-1}(-1),D^1(-1)]=T^0(-2),\quad &[D^{-1}(-1),S^1(0)]=D^0_1(1),\\ \nonumber
&[D^{-1}(-1),T^1_2(2)]=-\frac{1}{3}Q^0(-1)+\frac{4}{3} D^0_1(-1)\\ \nonumber
\end{align}

\par
%
%

\end{document}